\documentclass[prl,superscriptaddress,showpacs,keywords,reprint]{revtex4-1}
\usepackage{graphicx,amsmath,amssymb,amstext,amsfonts,color}

\usepackage[colorlinks,linkcolor=red,citecolor=blue,urlcolor=blue]{hyperref}

\newcommand{\be}{\begin{equation}}
\newcommand{\ee}{\end{equation}}
\newcommand{\ben}{\begin{eqnarray}}
\newcommand{\een}{\end{eqnarray}}
\newcommand{\bes}{\begin{subequations}}
\newcommand{\ees}{\end{subequations}}
\newcommand{\bF}{\begin{figure}}
\newcommand{\eF}{\end{figure}}

\def\tr{ {\rm{Tr }}\,}

\newcommand{\ket}[1]{\left|#1\right\rangle}
\newcommand{\bra}[1]{\left\langle#1\right|}

\newcommand{\mbf}[1]{\mathbf{#1}}

\begin{document}

\title{Information gain in tomography - A quantum signature of chaos}

\author{Vaibhav Madhok}
\email{vmadhok@gmail.com}
\affiliation{Center for Quantum Information and Control, University of New Mexico, Albuquerque, NM 87131-0001, USA}
\affiliation{Department of Physics and Computer Science, Wilfrid Laurier University, Waterloo, Ontario N2L 3C5, Canada}

\author{Carlos A. Riofr\'io}
\affiliation{Center for Quantum Information and Control, University of New Mexico, Albuquerque, NM 87131-0001, USA}
\affiliation{Dahlem Center for Complex Quantum Systems, Freie Universit\"{a}t Berlin, 14195 Berlin, Germany}

\author{Shohini Ghose}
\affiliation{Department of Physics and Computer Science, Wilfrid Laurier University, Waterloo, Ontario N2L 3C5, Canada}

\author{Ivan H. Deutsch}
\affiliation{Center for Quantum Information and Control, University of New Mexico, Albuquerque, NM 87131-0001, USA}

\date{\today}

\begin{abstract}
We find quantum signatures of classical chaos in various metrics of information gain in quantum tomography.  We employ a quantum state estimator based on weak collective measurements of an ensemble of identically prepared systems.  The  tomographic measurement record consists of a sequence of expectation values of a Hermitian operator that evolves under repeated application of the Floquet map of the quantum kicked top. We find an increase in information gain and hence higher fidelities in the reconstruction algorithm when the chaoticity parameter map increases.  The results are well predicted by random matrix theory.  
\end{abstract}
\maketitle

Signatures of chaos in quantum mechanics appear in a variety of contexts.  Examples include the level statistics associated with the random matrices of chaotic Hamiltonians~\cite{Bohigas, Haake},  hypersensitivity of dynamics to perturbations~\cite{per00, sc96}, the role of open quantum systems dynamics (decoherence/measurement) in the emergence of chaos~\cite{Zurek/Paz, bhj00}, and the connection between chaos and the dynamical generation of entanglement~\cite{Miller/Sarkar, Ghose, Wang, tmd08, Lakshminarayan/Bandyopadhyay2002}.  In this letter we identify and analyze a new signature that unifies the characterization of chaos in both classical and quantum physics --  chaos as a source for information gain in state estimation (tomography).  

At a fundamental level, chaos represents unpredictability, so this seems at odds with the goal of estimating an unknown state. On the flip side, however, this unpredictability represents the potential information to be gained in an estimation process. If everything is predicted and known, we learn nothing new.  The missing information in deterministic chaos is the {\em initial condition}.  In classical dynamics, a time history of a coarse grained trajectory at discrete times is an archive of information about the initial conditions given perfect knowledge about the dynamics.   Moreover, if the dynamics is chaotic, the rate at which we learn information increases due to the rapid Lyapunov divergence of distinguishable trajectories.  This information-theoretic picture is quantified by the Kolmogorov-Sinai (KS) entropy~\cite{s59}, which measures the information required to retrieve the increasingly fine-grained information about the initial condition that is necessary to maintain a constant coarse-grained prediction of the future chaotic trajectory.  

To probe the connection between quantum chaos and tomography, we consider a protocol based on the weak (nonprojective) collective measurement of an ensemble of $N_A$ identically prepared states that undergo well chosen dynamics~\cite{sjd05, rjd11}.  The time series of the measurement record provides the information used to reconstruct the initial condition.  The dynamics is ``informationally complete'' if the time history contains information about an arbitrary initial condition. Such a protocol has been implemented through continuous time measurement of an atomic spin ensemble driven by external magnetic fields while it is monitored by a weakly coupled laser probe~\cite{smith06, smith12}.   This protocol allows us to explore how the chaotic nature of the dynamics is revealed in the information content of the measurement record.  

Our goal in quantum tomography is to determine the state $\rho_0$ (here, for a spin $j$) given an ensemble prepared in the state $\rho_0^{\otimes N_A}$.  The system undergoes dynamics according to a prescribed unitary evolution, $U(t)$, and is measured collectively as described above.  The elements of the POVM, labeled by measurement outcomes $X(t)$ at time $t$, are taken to be~\cite{dj10}
\begin{equation}
E_{X(t)} = \frac{1}{\sqrt{2 \pi \sigma^2}}\exp\left\{-\frac{1}{2\sigma^2}\left(X(t) - J_z(t)\right)^2\right\}.
\label{POVM}
\end{equation}
We use the Heisenberg picture, so the observable being measured at time $t$ is $J_z(t) = U^\dag (t) J_z U(t)$, where $J_z$ is the $z$-projection of the collective angular momentum operator, $J_z = \sum_i j_z^{(i)}$, and $j_z^{(i)}$ is the projection for the $i^{th}$ spin.  The Gaussian spread in the POVM elements, $\sigma$, is set by the shot noise of the probe.  When the randomness of the measurement outcomes is dominated by the quantum noise in the probe rather than the uncertainty $\Delta J_z = \sqrt{N_A}~\Delta j_z$ (``projection noise''), quantum backaction is negligible, and the state remains approximately separable~\cite{dj10}.  In this case the stochastic measurement record, normalized by the number of atoms, is well approximated by
\begin{equation}
\label{record}
M(t) =X(t)/N_A = \tr(j_z(t) \rho_0) + W(t),
\end{equation}
where $W(t)$ is Gaussian white noise with spread $\sigma/N_A$.  

We parameterize the state in terms of a generalized Bloch vector $\mbf{r}$ for a $d$-dimensional Hilbert space ($d=2j+1$) by $\rho_0 = I/d + \sum_{\alpha=1}^{d^2-1} r_\alpha \Xi_\alpha$, where  $\{\Xi_\alpha \}$ form an orthonormal basis of traceless Hermitian operators.   Taking the record at discrete times $M_n =M(t_n)$, it then follows from Eq. (\ref{POVM}) that in the weak backaction limit, the probability of finding measurement history $\mbf{M}$ conditioned on the state $\mbf{r}$ is
\begin{eqnarray}
\label{eq:probabilitydist}
p\left(\mbf{M}| \mbf{r} \right) \propto \exp\left\{-\frac{N^2_A}{2 \sigma^2} \sum_i [ M_i - \sum_\alpha \mathcal{O}_{i \alpha} r_\alpha ]^2\right\}\\ \nonumber 
\propto\exp\left\{-\frac{N^2_A}{2 \sigma^2} \sum_{\alpha, \beta} \left(\mbf{r}-\mbf{r}_{ML} \right)_\alpha C^{-1}_{\alpha \beta} \left(\mbf{r}-\mbf{r}_{ML}\right)_\beta \right\},
\end{eqnarray}
where $\mathcal{O}_{n \alpha} = \tr\left(j_z(t_n) \,\Xi_\alpha\right)$ and $\mbf{C}^{-1} = \mbf{\mathcal{O}}^{T} \mbf{\mathcal{O}}$ is the inverse of the covariance matrix. The peak of this distribution is the (unconstrained) maximum-likelihood estimate of the Bloch vector, $\mbf{r}_{ML} = \mbf{C}\mbf{\mathcal{O}}^{T} \mbf{M}$. The problem of quantum tomography, therefore, reduces  to linear stochastic state estimation. The eigenvalues of $\mbf{C}^{-1}$ determine the relative signal-to-noise with which we have measured different observables (represented by its eigenvectors). When the covariance matrix is full rank, the measurement record is ``informationally complete.''  Due to finite signal-to-noise, however, the Bloch vector  $\mbf{r}_{ML}$ may not be associated with a physical density matrix with nonnegative eigenvalues, and therefore we must constrain the solution. The final estimate $\bar{\mbf{r}}$ is found as the closest matrix according to 
\begin{eqnarray}
\label{r_est}
\bar{\mbf{r}} = \arg \min \sum_i [ M_i - \sum_\alpha \mathcal{O}_{i \alpha} r_\alpha ]^2 , \\ \nonumber 
\text{ s.t., }\,  \frac{1}{d}I + \sum_{\alpha=1}^{d^2-1} \bar{r}_\alpha \Xi_\alpha \ge 0.
 \end{eqnarray}
We find the solution efficiently via convex optimization.

We now come to the central question of this work.  How is the information content in the measurement record related to the complexity of the dynamics encoded in $U(t)$?  To simplify the analysis, we consider periodic application of a given Floquet operator $U_{\tau}$, so that at the $n^{th}$ time step $U(n \tau) = U_{\tau}^n$. The measurement record generated by such periodic evolution is generally not informationally complete; it lacks information about a matrix subspace of dimension $\ge d - 2$ out of the total dimension $d^2 - 1$~\cite{mrfd10}.  
However, the condition that $\bar{\rho}$ be a positive matrix is a powerful constraint, that effectively allows ``compressed sensing''~\cite{gross10, smith12}, resulting in high fidelity reconstruction with the available information in the measurement record generated by the orbit of a single $U_{\tau}$.

Our goal is to relate the information generating power of $U_{\tau}$ to the properties of quantum state reconstruction according to the protocol above.  For this purpose, we consider a well-studied paradigm of quantum chaos: the kicked top~\cite{Haake}, described by the Floquet operator 
\begin{equation}
\label{Floquet}
U_{\tau} = e^{\frac{-i \lambda j_z^2}{2j}} e^{-i\alpha j_x}.    
\end{equation}
In our analysis, we fix $\alpha = 1.4$ and choose $\lambda$ to be our chaoticity parameter. As we vary $\lambda$ from $0$ to $7$, the classical limit of the dynamics change from highly regular to completely chaotic. In the quantum description, as the dynamics becomes globally chaotic, and for $j \gg 1$, the Floquet operator has the properties of a random matrix picked from the appropriate ensemble.  It is this randomness that leads to the analog of ergodic mixing for quantum systems. We take $j=10$ in our studies, which is sufficient to achieve the statistics of random matrices, but small enough to be essentially quantum.

\begin{figure*}
[t]\resizebox{18.1cm}{!}
{\includegraphics{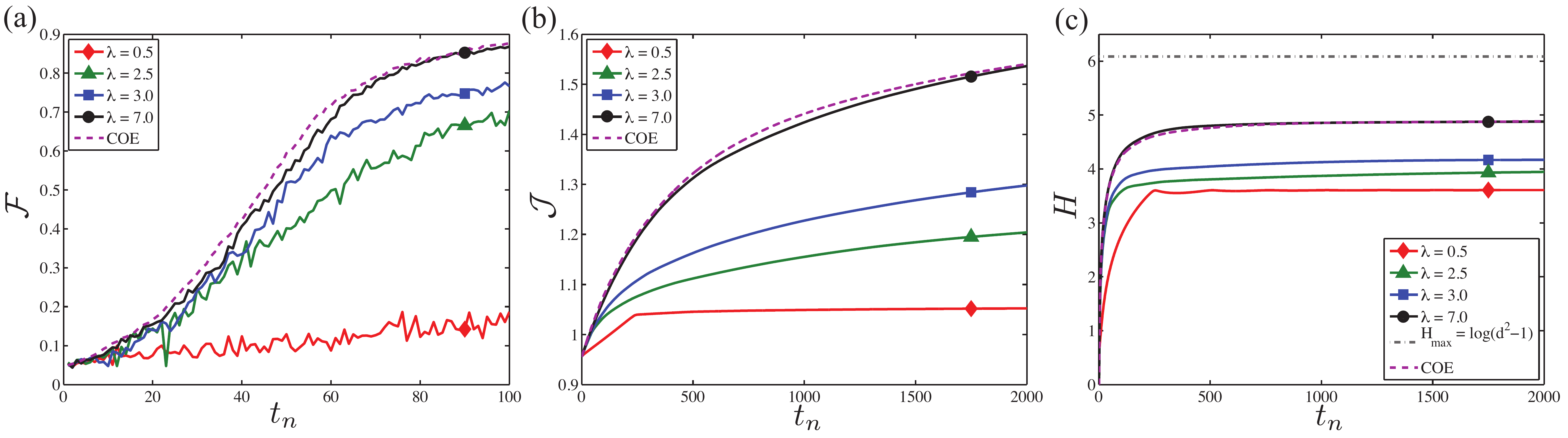}}
\caption{Information gain in tomography as quantified by various metrics, given a measurement record generated by iterations of a quantum-kicked-top Floquet map $U_{\tau} = \exp\{-i \lambda j_z^2/(2j)\} \exp\{-i\alpha j_x\}$,  for a spin $j=10$.  The value of $\alpha$ is fixed and $\lambda$ serves as the chaoticity parameter, varying from regular dynamics, $\lambda =0.5$, to fully chaotic, $\lambda=7.0$.  All results are shown as the average of 100 Haar-random pure states.  (a) Fidelity of state reconstruction. (b) The Fisher information of estimating the parameters that define the state.  (c) The Shannon entropy of the normalized eigenvalues of the inverse of covariance matrix of the likelihood function. The maximum possible Shannon entropy is shown as the dashed-dotted line.  In all cases, both the rate of growth and the final value of the information metric are increased with higher values of the chaoticity parameter, $\lambda$. In the fully chaotic regime, $\lambda =7.0$, the results are well predicted by tomography performed with a the measurement record generated by a Haar-random matrix picked from the circular orthogonal ensemble (COE).  The COE results, averaged over 100 random pure states, are plotted as the dashed line in (a)-(c).
}
\label{Fig1}
\end{figure*}

We study the behavior of our reconstruction algorithm for an ensemble of 100 random pure states sampled from the Haar measure on SU($d$), where $d=2j+1=21$.  The dynamical evolution of the measurement record is generated by repeated application of the kicked top Floquet operator in Eq. (\ref{Floquet}). Figure \ref{Fig1}a shows the fidelity of the state estimate, $\bar{\rho}$, relative to the true state, $\ket{\psi_0}$,  $\mathcal{F} =\bra{\psi_0}  \bar{\rho}  \ket{\psi_0}$, averaged over the ensemble of random states as a function of time, for different values of chaoticity in the kicked top. Two important features are apparent.  As the level of chaos increases, both the rate of increase of fidelity and its value after 100 periods of control by the Floquet operator increases.  When the classical description is globally chaotic, the corresponding quantum Floquet operator is well described as a random matrix and, together with the positivity constraint, allows us to accomplish high fidelity quantum state reconstruction.
We can further quantify the correlation between chaos and the performance of quantum state estimation using information theoretic metrics.  Brukner and Zeilinger~\cite{brukner99} defined the information available in a quantum measurement of a system, $E$, by the uncertainty of the outcomes summed over a set of mutually complementary experiments. \v{R}eh\'{a}\v{c}ek and Hradil showed that this uncertainty is equal to the  Hilbert-Schmidt distance between the true and estimated state in quantum tomography, averaged over many runs of the estimator, $E =\langle \tr\{(\rho_0 - \bar{\rho})^2\}\rangle $~\cite{rehacek02}, which can be expressed as the total uncertainty in the Bloch vector components, $E = \sum_\alpha\langle (\Delta r_\alpha)^2 \rangle$.  These uncertainties are always greater than the Cramer-Rao bound, $\langle (\Delta r_\alpha)^2 \rangle \ge \left[ \mbf{F}^{-1} \right]_{\alpha \alpha}$, where $\mbf{F}$ is the Fisher information matrix associated with the conditional probability distribution, Eq. (\ref{eq:probabilitydist}), and thus $E \ge \tr\, \mbf{F}^{-1}$.  In the limit of negligible quantum backaction, we saturate this bound, as our probability distribution is Gaussian, regardless of the state.  In that case, the Fisher information matrix equals the inverse of the covariance matrix, $\mbf{F} = \mbf{C}^{-1}$, in units of $N_A^2/\sigma^2$.  Thus, a metric for the total information gained in tomography is the inverse of this uncertainty,
\begin{equation}
\mathcal{J} = \frac{1}{\tr(\mbf{C})}  = \frac{1}{ \tr\left(\left( \mbf{\mathcal{O}}^T \mbf{\mathcal{O}} \right)^{-1}\right)},
\end{equation}
which measures the total Fisher information. Quantum tomography can also be viewed as a form of ``parameter estimation,''  i.e., estimate the Bloch vector components that define $\rho_0$.  The Fisher information, then, quantifies how well our estimator can predict these parameters from the data, regardless of the state.

In Fig. {\ref{Fig1}}b, we plot $\mathcal{J}$ as a function of time, generated by repeated application of the kicked top dynamics described above.  As before, we see the close correlation between the degree of chaoticity and the information gain in tomography.  Note that the inverse covariance matrix is never full rank in this protocol, and $\mathcal{J}$ is always ill defined.  We rectify this, regularizing $C^{-1}$ by adding to it a small fraction of the identity matrix, similar to the Tikhonov regularization (see,  e.g., \cite{boyd08}). In this way, our estimator ignores the directions in the space of observables that are largely unmeasured, and then makes the best guess consistent with the positivity constraint.  

There is a close relationship between Fisher Information and fidelity as a metric for information gain. When $\rho_0$ is a pure state, the average Hilbert-Schmidt distance, $E =1/\mathcal{J}= 1-\langle \tr\bar{\rho}^2\rangle -2 \langle \mathcal{F} \rangle$~\cite{hradil03}.  A correlation between chaos in the dynamics and the information gain as seen in the average fidelity implies that the Fisher information will exhibit the same correlation. Moreover, the Fisher information can be further related to a true information metric -- the mutual information $\mathcal{I}[\mbf{r};\mbf{M}]$ -- defined as the information we obtain about $\mbf{r}$ from measurement record $\mbf{M}$, which is given by $\mathcal{I}[\mbf{r};\mbf{M}] = H(\mbf{M}) - H(\mbf{M}|\mbf{r})$~ \cite{ct}.  Here, $H$ is the Shannon entropy of the given probability distribution.  Assuming perfect knowledge of the dynamics, the entropy of the measurement record, $H(\mbf{M})$, irrespective of the state, is due solely to shot noise, and is thus constant.  This is analogous, in the classical case, to the entropy associated with equal a priori probability to find a trajectory in one of the coarse grains of phase space.  Neglecting irrelevant constants, the mutual information between the Bloch vector and a given measurement record is then specified by the  entropy of the conditional probability distribution, Eq. (\ref{eq:probabilitydist}),  
\begin{equation}
\mathcal{I}[\mbf{r};\mbf{M}] = -H(\mbf{M}|\mbf{r}) = - \frac{1}{2}\log\left(\text{det}\mbf{C}\right) = \log (1/V),
\end{equation}
where $V$ is the volume of the error-ellipsoid whose semimajor axes are defined by the covariance matrix.

In order to maximize the information gain, we seek the dynamics that maximizes $1/V = \sqrt{\text{det }(\mbf{ C^{-1}})}$. An important constraint is that after time $t_n$,
\begin{equation}
\tr(\mbf{C}^{-1}) = \sum_{i,\alpha} \left( \mathcal{O}_{i,\alpha}\right)^2 = n \| \mathcal{O} (0)\|^2,
\end{equation}
where $\| \mathcal{O}(0) \|^2=\sum_\alpha \tr (\mathcal{O}(0) \Xi_\alpha)^2$ is the Euclidean square norm, with  $\mathcal{O}(0) =j_z$ for our example.   The right-hand-side of this equation is independent of $U$ and increases monotonically with time. It then follows from the inequality of arithmetic and geometric means, 
 \begin{equation}
 \text{det }(\mbf{ C^{-1}}) \le \left(\frac{1}{D} \tr(\mbf{C}^{-1}) \right)^D = \left(\frac{n}{D}\| \mathcal{O} (0)\|^{2}\right)^D  .
 \end{equation}
where $D =d^2-1$ is the rank of the regularized covariance matrix. The maximum possible value of the mutual information is attained when all eigenvalues are equal and the above inequality is saturated, implying that the error ellipsoid is a hypersphere.  At a given time step, the dynamics that gives the largest mutual information is the one that mixes the eigenvalues most evenly.  We quantify this by the Shannon entropy $H$ of the eigenvalues of the inverse of covariance matrix, $\lambda_\alpha$, normalized to as a probability distribution, $p_\alpha = \lambda_\alpha/\tr(\mbf{C^{-1}})$. Figure \ref{Fig1}c, shows $H(\vec{p})$ as a function of time generated by the kicked top Floquet map.  Again, we see a close relationship between the degree of chaoticity of the map and this metric for information gain in tomography.
This makes sense physically.  In order to extract the maximum information about a random state, we must measure all components of the Bloch vector with maximum precision.  Given finite time, we obtain the best estimate by dividing equally between all observables. 

Finally, we show that the information gain generated by the quantum-chaotic dynamics is fully consistent with random matrix theory.  The standard kicked top dynamics is time-reversal invariant without Kramer's degeneracy~\cite{Haake}.  For parameters in which the classical dynamics is globally chaotic, we thus expect the Floquet operator to have the statistical properties of a random matrix chosen from the circular orthogonal ensemble (COE)~\cite{Haake}. Figure \ref{Fig1} shows the behavior of the fidelity, the Fisher information, and Shannon entropy of the inverse of the covariance matrix as a function of the number of applications for a measurement record generated by a typical random unitary picked from the COE (dotted line). We see excellent agreement between our predictions from random matrix theory and the calculation for the evolution by the kicked-top Floquet map in completely chaotic regime, $\alpha=1.4, \lambda=7$.  Note that the maximum Shannon entropy, $H_{max}=\log(d^2-1)$, is never attained.  This is because our measurement record is generated by an orbit of repeated application of a single $U_\tau$.  Nonetheless, the measurement record, together with the positivity constraint, is sufficient to yield high fidelity reconstruction when the Floquet map is deep in the chaotic regime~\cite{mrfd10}.
 
In summary, complex dynamics reveals more information about the initial condition as one observes the system over the course of time.  Classically, chaotic dynamics, with its Lyapunov sensitivity to the initial conditions, generates an exponentially expanding archive of information about the initial state if we are able to track a trajectory with a constant coarse-grained resolution. Similarly, we found that the rate at which one obtains information about an initially unknown quantum state in quantum tomography is correlated with the extent of ``quantum chaos'' in the system. i.e., the degree to which the unitary dynamics maps a localized coherent state to a random state in Hilbert space.  Quantum tomography thus provides a forum in which to unify the the notions of complexity of chaotic dynamics in classical  and quantum worlds.  We quantified the information gain in a variety of metrics, including reconstruction fidelity, Fisher information, mutual information, and Shannon entropy. When the system is fully chaotic, the rate of information gain is well predicted by random matrix theory. This novel signature of chaos can be explored using current experimental techniques in the setting of cold atoms interacting with lasers and magnetic fields.

We thank Poul Jessen for useful discussions.  IHD, VM, and CR were supported by NSF Grant PHY-0903692, PHY-0903953, and PHY-1212445. This research was supported by the Natural Sciences and Engineering
Research Council of Canada and the Ontario Ministry of Economic
Development and Innovation.

\end{document}